\documentclass[twocolumn]{webofc}
\usepackage{graphicx}
\usepackage{amsmath}
\usepackage{dcolumn}
\usepackage{bm}
\usepackage{color}
\usepackage[varg]{txfonts}   
\usepackage{hyperref}
\usepackage{url}
\newcommand{\be}{\begin{equation}}
\newcommand{\ee}{\end{equation}}
\newcommand{\bea}{\begin{eqnarray}}
\newcommand{\eea}{\end{eqnarray}}

\hypersetup{colorlinks=true,citecolor=blue,urlcolor=blue,linkcolor=blue}
\begin{document}
\title{Novel microscopic approaches  for Spin-Isospin excitations and Beta-decay 
}
%
%

\author{\firstname{Hiroyuki} \lastname{Sagawa}\inst{1,2,3}\fnsep\thanks{\email{sagawa@ribf.riken.jp}} 
}

\institute{RIKEN  Nishina Center for Accelerator-based Science, Wako 351-0198, Japan
\and
     Mathematics and Physics, University of Aizu, Fukushima, Japan   
\and
Institute of Theoretical Physics, Chinese Academy of Sciences, Beijing 100190, China}


\abstract{
We explore   unsolved  nuclear structure problems related with  the spin and  isospin degree of freedom
by using microscopic models which accommodate realistic isoscalar and isovector pairing interactions,  and also tensor correlations. 
 For the attempt of universal theoretical framework for both nuclear and astrophysical phenomena, we adopt a self-consistent Hartree-Fock (HF)+random phase approximation (RPA) models, and  a state-of-the-art beyond mean field model, Subtracted Second RPA (SSRPA),  including the couplings to two-particle two-hole states. The quenching problems of magnetic dipole and Gamow-Teller  transitions are discussed in terms of the coupling to 2p-2h configurations and also the tensor correlations.  The $\beta$ decay life time of semi-magic and magic nuclei are  discussed in RPA and SSRPA models.  
}
\maketitle
\section{Introduction}
\label{intro}
The spin-isospin excitations including beta-decay may have strong  impacts on the study of strong interactions in nuclear medium, and also the astrophysical 
phenomena such as 
the r-process nucleosynthesis together with the photonuclear cross sections,  and  the  large-scale nucleosynthesis network calculations to create elements in the universe.   In addition, the double-beta decay processes are induced by Gamow-Teller (GT) and spin-dipole (SD) transitions from mother to grand daughter nuclei.  Especially, 
the  zero-neutrino double-beta decay is the key process to  obtain the information of neutrino-mass puzzle, and consequently may provide the evidence beyond the standard model of  elementary particles.   Thus, quantitative understanding of GT and SD excitations have not only strong impacts on the nuclear many-body problems associated with the  spin-isospin degree of freedom, but also the physics of interdisciplinary fields.

We explore   extensively new and old, but still unsolved,  nuclear structure problems induced by the spin and  isospin degree of freedom
by using microscopic models which accommodate realistic isoscalar and isovector pairing interactions,  and also tensor correlations \cite{Ha}. 
 For the attempt of universal theoretical framework for both nuclear and astrophysical phenomena, we adopt a self-consistent Hartree-Fock (HF)+random phase approximation (RPA) models embedded  the tensor interactions \cite{Cao}.   We further extend our approach to  a state-of-the-art beyond mean field model, Subtracted Second RPA (SSRPA) \cite{Yang} including the couplings to two-particle two-hole (2p-2h) states. 
Especially we study magnetic dipole (M1) \cite{Cao,Yang}, charge-exchange Gamow-Teller (GT)  and spin-dipole excitations. 
We mention also pigmy and giant resonances  induced by the tensor interaction.  

The quenching of the sum rule of GT excitations  has been a long-standing problem both experimentally and theoretically.  Our novel microscopic approach, SSRPA, may provide 
  a new insight on the quenching  of GT sum rule strengths without introducing any free  parameters in the  calculations.
The SSRPA model  is further applied to the $\beta$ decay half-lives of four semi-magic and magic nuclei, $^{34}\rm Si$, 
$^{68,78}\rm Ni$ and $^{132}\rm Sn$ for which the standard RPA model does not work \cite{Yang}.  We show  that 
 the inclusion of the 2p-2h configurations in SSRPA model
shifts low-lying Gamow-Teller (GT) states downwards, and 
 leads to an increase of the 
$\beta$ decay phase space, 
and consequently improve calculated $\beta$ decay half-lives dramatically close to the experimental observations.  
The effect of tensor interaction on the $\beta$ decay half-life in SSRPA model is also 
pointed out to  change largely the half-lives by about 
one to two orders of magnitude
with respect to the ones obtained without tensor force.  

This paper is organized as follows. Section 2.1  is devoted to basic formulas of density functional theory (DFT) and SSRPA model.   The subtraction method and the Okubo-Lee-Suzuki similarity transformation is also discussed in the Section 2.2.  The quenching problem of spin-isospin excitations is discussed in Section 3.   Summary is given in Section 4.
\section{Density Functional Theory (DFT) and Nuclear Many-body problems}
\label{sec-1}

\subsection{Kohn-Sham theory and Subtracted second RPA}
The  Density Functional Theory (DFT) was proposed in  1964, when a seminal paper by P. Hohenberg and W. Kohn  was published \cite{HK64}. In this work, two theorems have been proved. The first theorem
is that the total energy $E$ of a system of $N$ fermions can be written
in terms of their density $\rho({\bf r})$ only, as $E[\rho]$; the second theorem proves
that the exact total energy can be found  minimizing the energy
as a function of  $\rho({\bf r})$;
\begin{equation}
E_0 = {\rm min}_{\rho} E[\rho],  
\end{equation}
where 
$E[\rho]$ is called Energy Density Functional (EDF).
Following this theorem, one does not need to use the total wave function  $\Psi(\vec r_1 \ldots \vec r_N)$ to calculate the exact energy.   This is a tremendous step forward since the total wave function is not only hard or impossible to calculate when $N$ is large and easier to obtain.
The density is instead a function of three coordinates, regardless of $N$. The Hohenberg-Kohn (HK) theorems also imply that the wave function itself is uniquely determined once $\rho$ is given and, thus, all observable property of the system can be  uniquely associated with the density.

The popularity of DFT stems from its conceptual elegance and ability to be accurate at a relatively low computational cost.
I should also tress the aspect of "universality", that is, one can apply the DFT model for a wide variety  of many-body systems from small and extremely large number of particles with one optimized EDF.  
Very often, the so-called Kohn-Sham (KS)  approach\index{Kohn-Sham (KS)} is adopted 
to implement the HK theorem into many-body problems \cite{Kohn-Sham65}. In this approach, the assumption is  that the density $\rho$ can be 
expressed in terms of auxiliary single-particle orbitals $\phi_i({\vec r})$:
\begin{equation}\label{eq:KS-basic}
\rho({\vec r}) = \sum_i \vert \phi_i({\vec r}) \vert^2.
\end{equation}
The kinetic energy is then given by  a formula 
\begin{equation}
T = \sum_i -\frac{\hbar^2}{2m} \int d^3r\ \phi_i^*({\vec r})
\nabla^2 \phi_i({\vec r}).
\end{equation}

In  the nuclear 
many-body problem, we have to introduce the spin and isospin coordinates, $s_i$ and $t_i$. 
The total wave function is expressed as $\Psi(x_1, x_2, \cdots, x_A)$ with $x_i\equiv(\vec{r}_i,s_i,t_i)$.
The nuclear Hamiltonian  has a form with two-body and three-body interactions,
\begin{equation}\label{eq:H_nucl}
H = \sum_{i=1}^A -\frac{\hbar^2}{2m}\nabla_i^2 + \frac{1}{2}\sum_{i \ne j = 1}^A
V_2(x_i,x_j) + \frac{1}{6} 
\sum_{\substack{i,j,k=1 \\ i\ne j, j\ne k, i\ne k}}^A
V_3(x_i,x_j,x_k) 
\end{equation}
where 
$m$ is the nucleon mass. 
In principle, 
four-body forces cannot  be ruled out, and it appears as $N^3L$ order diagram in the chiral effective field theory (ChEFT).    
The total energy is then expressed as an universal function of one-body density;
\begin{equation}
E(\rho)=\langle \Psi|H|\Psi \rangle=\langle \Psi|T+V|\Psi \rangle.  
\end{equation}
In the KS approach, the density is calculated based on the so-called "self-consistent" mean field approach.  That is, the one-body hamiltonian is derived by a functional derivative of EDF with respect to the density $\rho$,
\be  \label{KS1}
h({\bf r})=\frac{\delta E(\rho)}{\delta \rho}.
\ee
In the next step, the single-particle wave function is calculated by using the hamiltonian $h({\bf r})$;
\be \label{KS2}
h({\bf r}) \phi_i({\bf r})=\epsilon_i  \phi_i({\bf r}),
\ee
where $\epsilon_i $ and  $\psi_i({\bf r})$ are the eigenvalue and the eigenfunction of the hamiltonian, respectively. Then the density can be constructed as
\be \label{KS3}
\rho({\bf r})=\sum_{i=1}^A|\phi_i({\bf r})|^2.
\ee
The process from \eqref{KS1} to  \eqref{KS3}  is repeated starting from the first trial density $\rho_0({\bf r})$ until the solution is converged by certain criterion.  The repeating of process is called the self-consistent method to obtain the mean field density.

DFT was designed to describe the ground state properties such as the binding energies and the radii of many-body systems.  The model can be extended to describe the excited states.  In a small amplitude limit of time-dependent HF theory, one can obtain the model, so called, the random phase approximation (RPA).  Equivalently, the same RPA equation can be obtained by the qui-boson equation of motion model.  We briefly present main formalism of RPA  and subtracted second RPA (SSRPA) models \cite{MJ2022}.
In RPA and SRPA models, the excitation operator $Q_\nu^\dagger$ can be written as, 
\begin{equation}\label{eq1}
\begin{split}
	Q_\nu^\dagger&=\sum_{ph}(X_{ph}^\nu a_p^\dagger a_h-Y_{ph}^\nu a_h^\dagger a_p) \\
	&+\sum_{\substack{p_1<p_2\\h_1<h_2}}(X_{p_1p_2h_1h_2}^\nu a_{p_1}^\dagger a_{p_2}^\dagger a_{h_2}a_{h_1}
	-Y_{p_1p_2h_1h_2}^\nu a_{h_1}^\dagger a_{h_2}^\dagger a_{p_2}a_{p_1}),  
	\end{split}
\end{equation}
where the first line correspond to 1p-1h excitations in the RPA model and the second line is 2p-2h excitations included in the SRPA model.
The footnotes $p, p_1$, $p_2$ denote particle states, while $h,  h_1$, $h_2$ are hole states. $X$'s and $Y$'s are forward and backward amplitudes. The RPA and SRPA equations have the same form as
\begin{equation}\label{eq2}
\begin{bmatrix}
A&B\\-B^*&-A^*
\end{bmatrix}	
\begin{bmatrix}
X^\nu \\ Y^\nu
\end{bmatrix}
=\hbar\omega_\nu\begin{bmatrix}
X^\nu\\Y^\nu
\end{bmatrix}, 
\end{equation}
where the SRPA has the double column structure; $A$ and $B$ are $2\times2$ matrices and $X$ and $Y$ are two column vectors,
\begin{equation}\label{eq3a}
\begin{split}
&A=
\begin{pmatrix}
	A_{11}&A_{12}\\A_{21}&A_{22}
\end{pmatrix}
, B=
\begin{pmatrix}
B_{11}&B_{12}\\B_{21}&B_{22}
\end{pmatrix},\\
&X=
\begin{pmatrix}
X_1^\nu\\X_2^\nu
\end{pmatrix}
,Y=
\begin{pmatrix}
Y_1^\nu\\Y_2^\nu
\end{pmatrix}
,\end{split}
\end{equation}
while RPA has single column structure.  
The indices 1 and 2 are a shorthand notation for the 1p-1h and 2p-2h configurations, respectively. 
The matrices $A_{12}$ , $B_{12}$ denote the coupling of 1p-1h with 2p-2h configurations, 
and $A_{22}$, $B_{22}$ denote the coupling of 2p2h configurations among themselves. Simply speaking, 2p-2h configurations are completely omitted in the RPA model.  
Diagrammatically , these couplings are shown in Figs. 1(a), 1(b) and 1(c).
\begin{figure}[t]
\centering
\includegraphics[width=8.8cm,clip,bb=20 0 700 500]{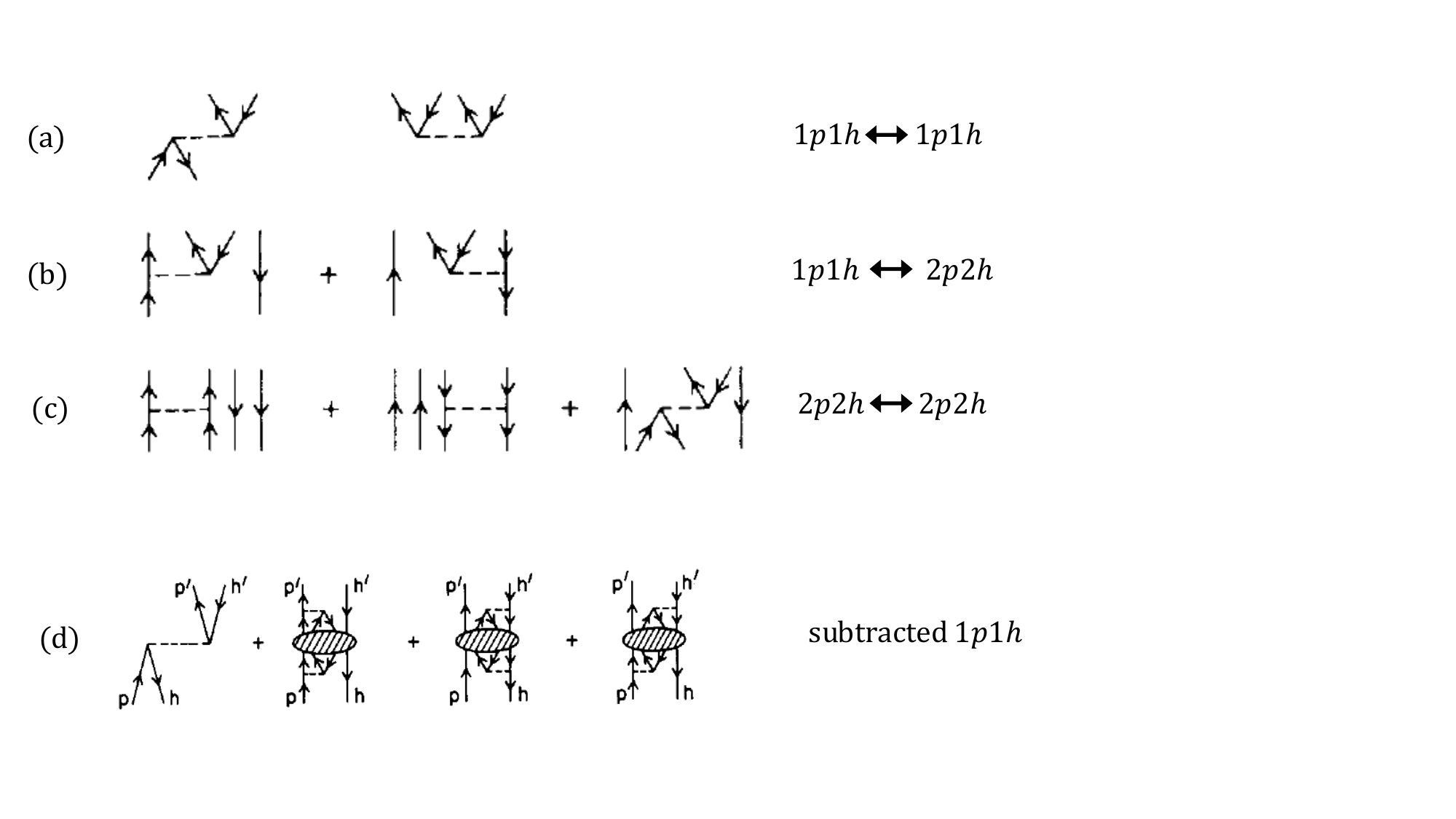}
\vspace{-1cm}
\caption{Diagrammatic representations of SRPA matrix elements. (a)  between 1p-1h and 1p-1h configuration. (b) between 1p-1h and 2p-2h configuration (c) between 2p-2h and 2p-2h configuration (d) subtracted 1p-1h matrix element. 
See the text for details. }
\label{fig-2}       
\end{figure}
The matrix elements of Eq. \eqref{eq3a} are equal to
\begin{eqnarray}
\hspace{-1cm}	A_{11}&=&A_{ph;p^\prime h^\prime}
	=\langle HF|[a_h^\dagger a_p,[H,a_{p^\prime}^\dagger a_{h^\prime}]]|HF\rangle   \nonumber \\
	&=&(E_p-E_h)\delta_{pp^\prime}\delta_{hh^\prime}+\bar{V}_{ph^\prime hp^\prime},  \label{eq4}\\
B_{11}&=&B_{ph;p^\prime h^\prime}   \nonumber \\
&=&-\langle HF|[a_h^\dagger a_p,[H,a_{h^\prime}^\dagger a_{p^\prime}]]|HF\rangle    
=\bar{V}_{pp^\prime hh^\prime},  \label{eq5}\\
A_{12}&=&A_{ph;p_1p_2h_1h_2}     \nonumber \\
&=&\langle HF|[a_h^\dagger a_p,[H,a_{p_1}^\dagger a_{p_2}^\dagger a_{h_2}a_{h_1}]]|HF\rangle  \label{eq6}, \\
A_{22}&=&A_{p_1p_2h_1h_2;p_1^\prime p_2^\prime h_1^\prime h_2^\prime}   \nonumber \\
&=&\langle HF|[a_{h_1}^\dagger a_{h_2}^\dagger a_{p_2}a_{p_1},[H,a_{p_1^\prime}^\dagger a_{p_2^\prime}^\dagger a_{h_2^\prime}a_{h_1^\prime}]]|HF\rangle ,  \label{eq7}  \nonumber \\
\end{eqnarray} 
where $E_p$ and  $E_h$ are the HF particle and hole energies,  respectively, 
 and $\bar{V}$ is the anti-symmetrized matrix element of 
residual interaction. The quasi boson approximation (QBA) is used in the derivation of matrix elements, and it turns out that 
$B_{12}$, $B_{21}$ and $B_{22}$ have no contributions to the SRPA matrices:
\begin{equation}
B_{12}=B_{21}=B_{22}=0.
\end{equation} 

It is well-known that the SRPA has a convergence problem;  {a larger model space makes a lower energy 
spectra  {than the} physical energy region.}
To cure this problem, the subtraction method was proposed.  This procedure will start from the convolution of $2p$-$2h$ model space into the $1p$-$1h$ model space.  Then we get the energy dependent matrices $A$ and $B$ for $1p$-$1h$ model space as 
\begin{eqnarray}
\small
  \begin{split}\label{AB-sub}
  A_{11^\prime}(\omega)=
  A_{11^\prime}+&\sum_{2}A_{12}(\omega+i\eta- A_{22})^{-1}A_{2 1^\prime}  \nonumber  \\
  +&\sum_{2}B_{12}(\omega+i\eta- A_{22})^{-1}B_{21^\prime}, 
  \\
  B_{11^\prime}(\omega)=
  B_{11^\prime}+&\sum_{2}A_{12}(\omega+i\eta-A_{22})^{-1}B_{21^\prime}  \nonumber  \\
  +&\sum_{2}B_{12}(\omega+i\eta-A_{22})^{-1}A_{2 1^\prime} .
  \end{split}
\end{eqnarray}
{Two additional terms are $\omega$-dependent and will induce  non-convergence of SRPA because $A_{11^\prime}(\omega)$ and $B_{11^\prime}(\omega)$ for $1p$-$1h$ model space 
depend on the size of $2p$-$2h$ model space and consequently  the SRPA solutions also depend on the model space.}   To cure this problem, the $A$ and $B$ matrices in the SRPA \eqref{eq2} 
 are modified. as
  \begin{equation}\label{eq9}
  \small
  \begin{split}
  A_{11^\prime}^S=&
  A_{11^\prime}+\sum_{2}A_{12}(A_{22})^{-1}A_{2 1^\prime}+\sum_{2}B_{12}(A_{22})^{-1}B_{21^\prime}
  \\
  B_{11^\prime}^S=&
  B_{11^\prime}+\sum_{2}A_{12}(A_{22})^{-1}B_{21^\prime}+\sum_{2}B_{12}(A_{22})^{-1}A_{2 1^\prime} 
  \end{split}
  \end{equation}
  Notice that  the extra terms in Eq. \eqref{eq9} cancel exactly with the additional terms in Eq. \eqref{AB-sub} in the static limit $\omega\rightarrow 0$.
  {This subtraction method is unavoidable since EDF was constructed originally to reproduce nuclear structure observables in the mean field levels such as HF and RPA.}
  
\begin{figure}[t]
\centering
\includegraphics[width=8cm,clip,bb=0 0 450 350]{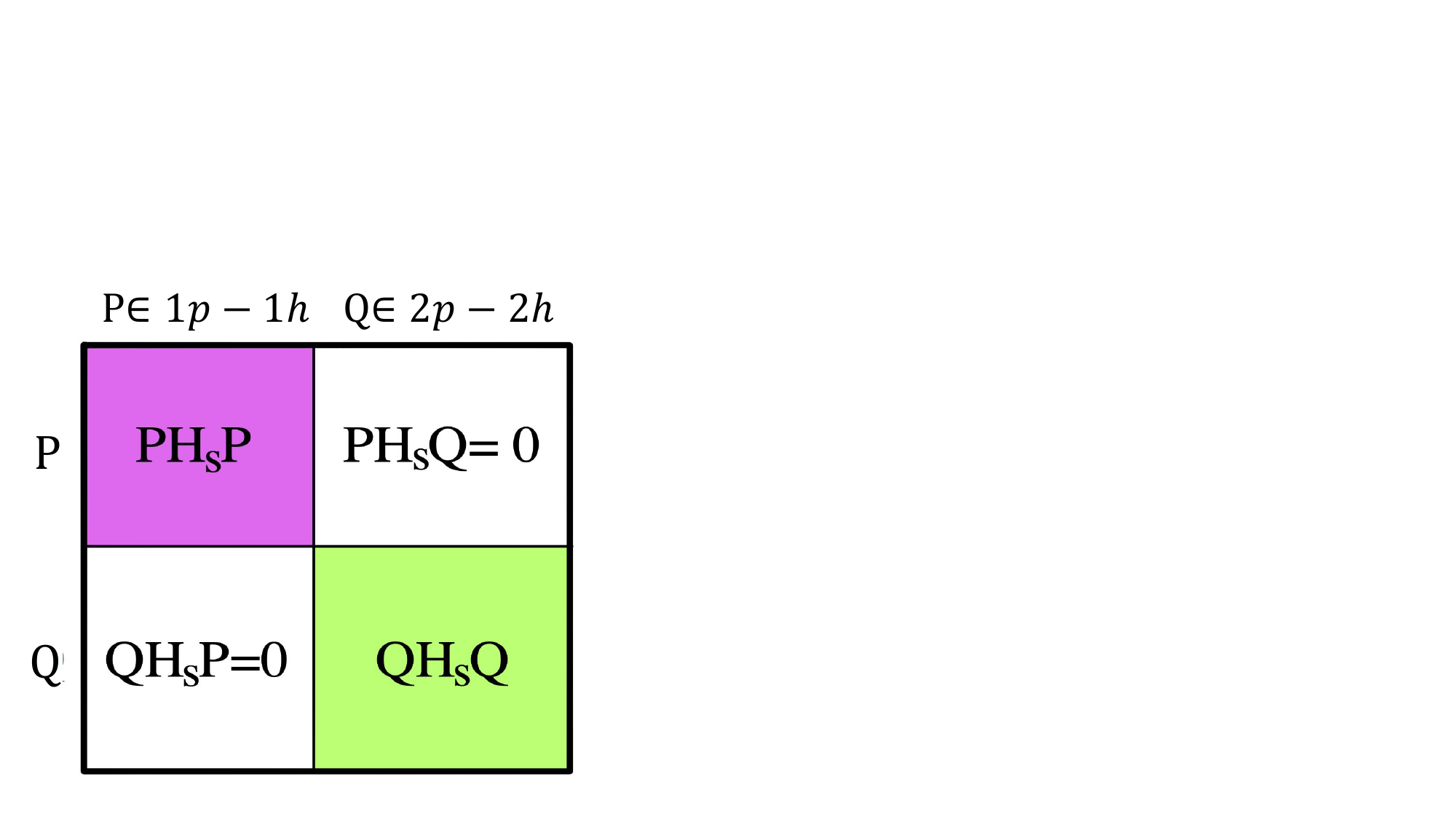}
\caption{Diagrammatic representations of Lee-Suzuki similarity transformation.
$P$ space represents 1p-1h model space, while $Q$ represents 2p-2h model space.}
\label{fig-1}       
\end{figure}

\subsection{ Okubo-Lee-Suzuki similarity transformation and SSRPA}
The coupled cluster model with single- and doublet pairs (CCSD) is a similar approach to SRPA to include $2p$-$2h$ configurations on top of $1p$-$1h$ configurations.  The essential difference is that the interaction is renormalized according to the size of model space by using 
a similarity transformation (it is called " Okubo-Lee-Suzuki" transformation ) or the similarity renormalization group (SRG) technique \cite{Suzuki1980}.
According to the  Okubo-Lee-Suzuki method,  we start from a many-body system which is described by the Schr\"odinger equation
\begin{equation} 
H |\Psi\rangle = E |\Psi\rangle,
\end{equation} 
with an ab initio Hamiltonian
\begin{equation} 
H=H_0+V,
\end{equation} 
where $H_0$ is the one-body Hamiltonian and $V$ is the residual interaction, and $|\Psi\rangle$ is the exact wave function and $E$ is the exact eigenvalue.   In standard interacting shell models, the  full model space is separated into two parts: the active space denoted $P$ space, and the inactive space called $Q$ space. We define the projection operators $P$ and $Q$, which project out each space,  and satisfy
\begin{equation} 
P+Q=1, \hspace{0.5cm} PQ=0.
\end{equation} 
In the mean field model, we define that the $P$ space is $1p$-$1h$ model space and $Q$ space is $mp$-$mh$ states with $m=2,3,\cdots$.   
The operators $P$ and $Q$ satisfy the commutation relations with $H_0$; 
\begin{equation} 
[P, H_0]=[Q,  H_0]=0,
\end{equation} 
and called eigenprojectors.  Consequently, we can obtain
\begin{equation} 
PH_0Q=QH_0P=0.
\end{equation} 

The objective of the  Okubo -Lee-Suzuki is to construct,  from the full Hamiltonian $H$,  an effective Hamiltonian $H_{eff}$, which acts only
in the $P$ space and satisfies the condition that any eigenvalue of $H_{eff}$ should
be one of the exact eigenvalues of the full Hamiltonian $H$. A general equation for
determining $H_{eff}$ can be derived by the use of the similarity transformation theory.
We consider a similarity transformation of the Hamiltonian $H$
\begin{equation} 
\mathcal{H_S}=X^{-1}HX,
\end{equation} 
 where $X$ is a transformation operator which is defined in  the entire Hilbert space.  {The operator $X$ is not unitary, but 
has its inverse $X^{-1}$.}  The transformed Hamiltonian $\mathcal{H_S}$ is decomposed into four terms
\begin{equation}  \label{PQ}
\mathcal{H_S}=(P+Q)\mathcal{H_S}(P+Q)=P\mathcal{H_S}P+Q\mathcal{H_S}P+P\mathcal{H_S}Q+Q\mathcal{H_S}Q.
\end{equation} 
To drive the effective Hamiltonian,
 \begin{equation} \label{Heff}
H_{eff}=P\mathcal{H_S}P=P(X^{-1}HX)P
\end{equation} 
from Eq. \eqref{PQ},  we require a condition
\begin{equation}  \label{PQ1}
Q\mathcal{H_S}P=Q(X^{-1}HX)P=0.
\end{equation} 
 Equation \eqref{PQ1} provides the necessary and sufficient condition for the determination
of $H_{eff}$, which has the eigenstate in $P$ space corresponding to  one of the exact eigenstate of the full Hamiltonian $H$.
 We can prove this statement by the similarity transformation
theory in the eigenvalue problem. If $X$ is a solution of
Eq.  \eqref{PQ1}, $H_{eff}$ in Eq.  \eqref{Heff} satisfies for any eigenstate $\phi$ in the $P$ space the following equation, 
\begin{eqnarray}  \label{H-eff}
H_{eff}|\phi\rangle&=&(P\mathcal{H_S}P+Q\mathcal{H_S}P)|\phi\rangle  \nonumber \\
  &=&\mathcal{H_S}P|\phi\rangle=\mathcal{H_S}|\phi\rangle=E_\phi|\phi\rangle,
\end{eqnarray} 
with the eigenvalue $E_\phi$ 
{so that $H_{eff}$ is equivalent to $\mathcal{H_S}$ in $P$ space.}   Therefore, the eigenvalue of $H_{eff}$ becomes also
the eigenvalue of $H$:
\be
X\mathcal{H_S}|\phi\rangle=HX|\phi\rangle, 
\ee
and
\be
X\mathcal{H_S}|\phi\rangle=XE_\phi|\phi\rangle =E_\phi X|\phi\rangle,
\ee
so that
\be\label{H-t}
H|\tilde {\phi}\rangle=E_\phi |\tilde {\phi}\rangle,
\ee
with $|\tilde {\phi}\rangle\equiv X|\phi\rangle$.  
Equations \eqref{H-eff} and \eqref{H-t} prove  that the eigenvalue of $H_{eff}$ agrees with one of
the eigenvalues of the full Hamiltonian $H$  because the similarity transformation does
not change the eigenvalues.  

The operator $X$ is conveniently expressed as
\begin{equation} \label{X1}
X=e^\omega,
\end{equation}   
where the operator $\omega$ has properties
\begin{equation} \label{X2}
\omega=Q\omega P\neq 0,  \,\,\,\, P\omega P=Q\omega Q=P\omega Q=0,
\end{equation} 
since $\omega$ is introduced to act a transformation from $P$-space to $Q$-space.  From the properties \eqref{X2}, Eq. \eqref{X1} is simplified as
\begin{equation} 
X=1+\omega,
\end{equation} 
since $\omega^2=\omega^3=\cdots=0$.

The effective Hamiltonian is rewritten to be
\begin{equation} 
H_{eff}=P\mathcal{H_S}P=PHP+PVQ\omega,
\end{equation} 
by using the properties of operator $\omega$.   Then the effective interaction is given by
\begin{equation} 
V_{eff}=H_{eff}-PH_0P=PVP+PVQ\omega.
\end{equation} 
The Hamiltonian $H_{eff}$ has  the eigenvalue $E_i$ and the corresponding eigenstate $|\phi_i\rangle$ as,  
\begin{equation} 
H_{eff}|\phi_i\rangle=(PH_0P+V_{eff})|\phi_i\rangle=E_i|\phi_i\rangle,
\end{equation} 
and the operator $\omega $ is expressed as  \cite{Suzuki1980}
\begin{equation} 
\omega =\sum_{i}^{d}\omega(E_i)|\phi_i\rangle \langle \phi_i |,  \,\,\, {\rm with}  \,\,\,\omega(E_i)=\frac{1}{E_i-QHQ}QVP,
\end{equation} 
where $d$ is the dimension of $P$-space.  Eventually the effective interaction of $P$ space is expressed as
\begin{equation}  \label{V-eff}
V_{eff}=PVP+\sum_{i}^{d}PVQ\frac{1}{E_i-QHQ}QVP|\phi_i\rangle \langle \phi_i |.
\end{equation} 
The eigenstate of full model space is also expressed as
\begin{equation} 
|\Psi_i\rangle=e^\omega|\phi_i \rangle=|\phi_i \rangle+\omega(E_i)|\phi_i \rangle.
\end{equation} 
Notice that the eigenenergy of $p$-space $E_i$ is the same as that of full space eigenstate $|\Psi_i\rangle$.

The similarity transformation provides the renormalization term in $V_{eff}$ in Eq. \eqref{V-eff}, which is exactly the same functional form as the subtraction term in SSRPA.
Since EDF and the effective Lagrangian include the renormalization term from the beginning, it is physically justified to subtract the corresponding term if one expand the model space from $P$-space to $P$+$Q$ space.  The adiabatic limit $\omega \rightarrow 0$ should be also reasonable since EDF is optimized for the data set of nuclear ground state and nuclear matter.

\subsection{Objectives and microscopic models}
\label{sec-2}
The spin-isospin excitations have been discussed in terms of collectivity, the quenching of sum rule strength, and the relations with elementally particle physics and astrophysics. The main interests are listed as, 
\begin{itemize}
\item 
Collectivity of spin and isospin excitations.
\item Landau parameters of spin and spin-isospin channels $g$ and $g^{\prime}$,  which may imply the  possible pion condensations at  high nuclear density. 
\item SU(4) symmetry in light nuclei and the role of isoscalar (T=0) pairing. 
\item Damping mechanism due to the  {configuration mixing.}
\item Effects of tensor correlations on M1 and GT excitations.
\item  Isobaric analogue state (IAS) and  Charge symmetry breaking (CSB) and charge independence breaking (CIB) interactions.
\item Super-allowed Fermi transitions and its implication on  Cabibbo-Kobayashi-Maskawa  (CKM) unitarity matrix, which has been studied in the context of  beyond standard model physics of elementary particles.  
\item Two-neutrino and neutrino-less double-beta decay, and neutrino mass puzzle.  
\end{itemize}

 Theoretically,  the interacting shell model and RPA are the models
which are widely applied to study the spin-isospin excitations. 
The shell models are developed in various versions, such as the large-scale interacting shell model with realistic and phenomenological interactions, the ab initio no core shell model, the ab initio 
coupled cluster model,  the Gamow shell model,  the ab initio 
Green's function Monte Carlo (GFMC) shell model, and the Monte Carlo shell model (MCSM) with intrinsic deformed Slater determinants.
More recently, the nuclear Lattice effective field theory is developed and applied to light and heavy nuclei.  
 The RPA models have also many variations  including self-consistent RPA based on the effective energy density functionals (EDFs), 
and conventional  RPA models based on the realistic or Landau-Migdal interactions. 
There are also models beyond mean field approach: 
\begin{itemize}
\item Particle-vibration coupling model (PVC),  Quasi-particle PVC (QPVC).
\item Relativistic quasiparticle time blocking approximation (RQTBA).
\item Coupled cluster model with singlet  and doublet clusters (CCSD).
\item Second RPA (SRPA), Subtracted SRPA (SSRPA) .
\item Self-consistent Green's function approach (SCGFA).
\item Generator coordinate method (GCM).
\end{itemize}

\section{Quenching spin-dependent operator}
The quenching problem of spin-dependent operator has been well recognized for many years.  They are  
\begin{itemize}
\item Magnetic moments   and M1 transitions represented by the operator       $g_s {\bm s}+ g_l{\bm l}$, 
where $g_s$ and $g_l$ are the spin and orbital $g$ factors, respectively.
\item IV Magnetic dipole transitions  represented by the operator   $g_s^{IV} {\bm \sigma}\tau_z$,  where $g_s^{IV}=(g_s^{n}-g_s^{p})/2$.
\item Gamow-Teller charge-exchange transitions    represented by the operator   $g_A  {\bm \sigma} t_{\pm}$, where $g_A$ is the axial vector coupling constant. 
\item  Spin-dipole charge exchange transitions represented by the operator  $g_A{\bm \sigma} Y_{1\mu} t_{\pm}$,   where  $Y_{1\mu}$ is the spherical harmonics. 
\end{itemize} 
A famous problem is the quenching of magnetic moment.  There have been discussed many times what is the origin of these quenching phenomena.
Arima and Horie pointed out that the configuration mixing is the most important effect on the quenching of magnetic moments.  Other effects such as $\Delta$-hole coupling and meson-exchange currents are also pointed out as
important mechanisms to induce the quenching of spin dependent matrix elements.  These effects revive in modern calculations of GT matrix elements: the configuration mixing is implemented in the large scale shell model calculations, coupled
cluster models and also SSRPA calculations.  The two-body currents including the pair  currents and pion-exchange currents are also discussed in  the framework of chiral effective field theory.

\begin{figure*}[t]
\centering
\includegraphics[width=17cm,clip,bb=0 0 900 470]{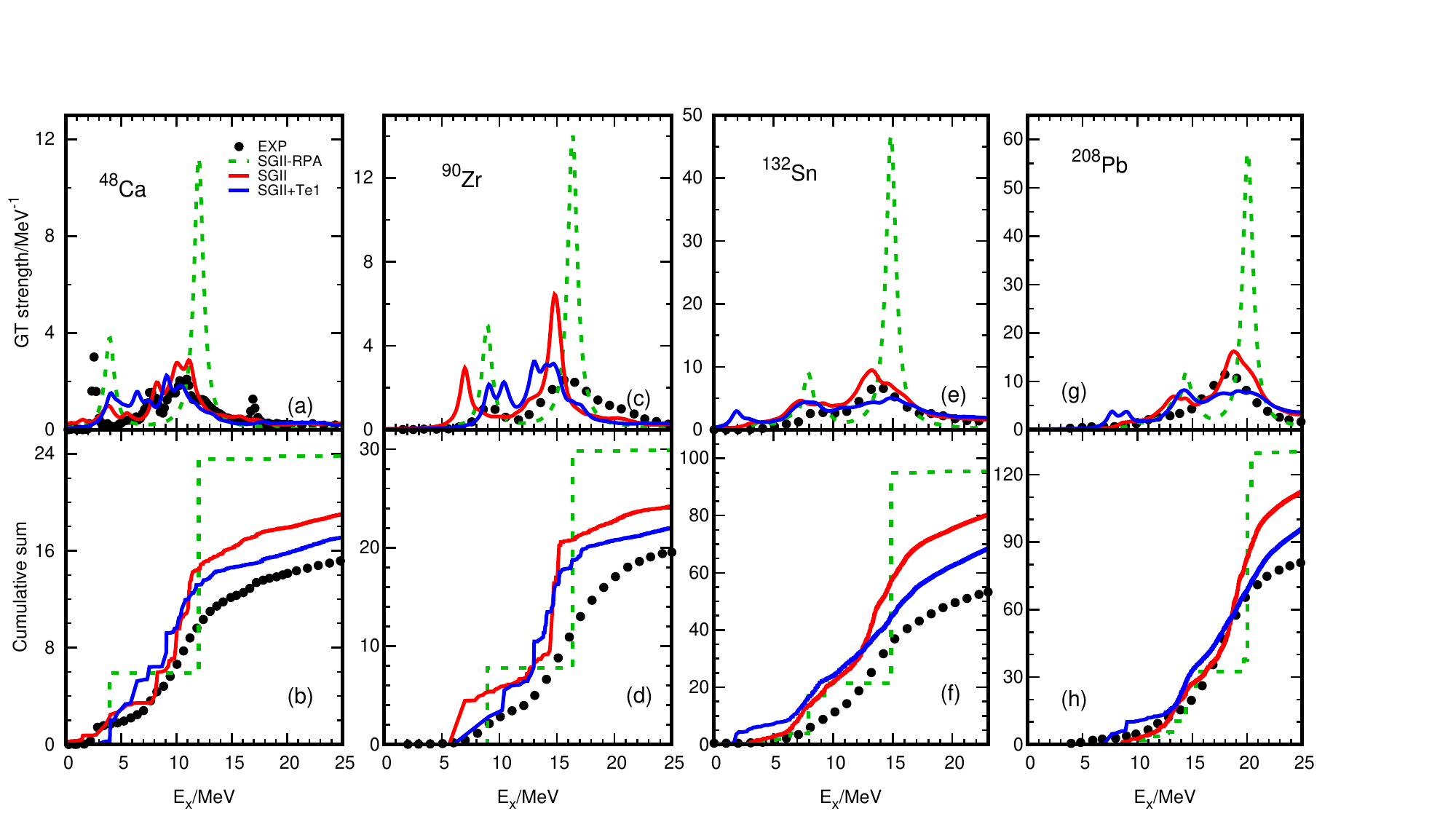} 
\caption{(upper panels) GT strength distributions of $^{48}$Ca, $^{90}$Zr, $^{132}$Sn and $^{208}$Pb calculated by RPA and SSRPA models. (lower panels) Cumulated sum of the GT strength.  Green dashed lines show RPA results, while 
red and blue solid curves correspond to SSRPA results without/ with tensor terms.  Experimental data are taken from Ref. \cite{GT-48Ca}  for $^{48}$Ca, Ref. \cite{GT-90Zr} for $^{90}$Zr, Ref. \cite{GT-132Sn} for $^{132}$Sn and 
Ref. \cite{GT-208Pb} for $^{208}$Pb, respectively.}
\label{fig:GT-SSRPA}       
\end{figure*}

We performed SSRPA calculations to study the quenching effect due to the 2-particle-2-hole configuration mixings on the GT excitations.  Results are shown in Fig. \ref{fig:GT-SSRPA}.  RPA calculations reproduce the peak positions of observed GT resonances in all nuclei, but the strengths are much larger than the experimental observations. 
As is seen in this figure, the Ikeda sum rule
\begin{equation}
S_- -S_+=\sum_n |\langle n|\sigma t_-|0\rangle|^2 -\sum_n|\langle n|\sigma t_+|0\rangle|^2=3(N-Z), 
\end{equation}
is exhausted  by 100\% in RPA calculations for all nuclei, while the experimental data exhaust only about 60\% of the Ikeda sum rule.  On the other hand,  the SSRPA calculations show about 20\% of the GT strength are shifted to higher energy region  than 20 MeV excitation energy.  The calculated GT strength around the resonance becomes even closer to the experimental finding when the tensor terms are included in the Skyrme EDF.
%
%

\subsection{Magnetic dipole (M1) transition in $^{48}$Ca}
We  investigate the role of tensor interactions on the M1 transitions in $^{48}$Ca \cite{MJ2024}. 
{The M1 operator reads}
\begin{equation} 
\hat{O}(M1)=\sum_i (g_s(i){\bf s}_i+g_l(i){\bf l}_i),
\end{equation} 
{where $g_s(i)\,\, (g_l(i))$ is the spin (orbital) $g$ factor given by $g_s(p)=5.586\,\, (g_l(p)=1.0)$ for protons and $g_s(n)=-3.826 \,\,(g_l(n)=0.0)$ for neutrons, respectively, in unit of nuclear magneton $\mu_N$.}
  We tabulate the unperturbed $p$-$h$ and RPA M1 excitation energies of SGII-T and 
SAMi-T without and with the tensor interactions  in Table 1. {The tensor part of SAMi-T was determined guided by ab initio relativistic Brueckner-Hartree-Fock (RBHF) studies on neutron-proton drops \cite{SAMi-T}, while that of SGII-T was optimized to reproduce excitation energies of GT and spin-dipole states of doubly-close shell nuclei \cite{Bai2011}}
  The dominant $p$-$h$ configuration is  $(1f_{7/2}\rightarrow 1f_{5/2})_\nu$ 
 for $^{48}$Ca   so that the orbital operator has no contribution to the M1 strength.
In the HF level, the triplet-odd term $U$ acts uniquely on the energy splitting of neutron spin-orbit partners \cite{Colo2007}.
 The large negative $U$ value increases the spin-orbit splitting of like-particles so that the tensor interaction of SGII+T with a larger negative $U$  gives much larger spin-orbit splitting compared with that of SAMi-T %
 as is seen 
 Table 1.  The strong triplet-odd term $U$ acts also largely on the  RPA correlations pushing upwards the excitation energies of M1 states. 

\begin{table*}[tb]
  \centering
  \caption{The excitation energy of $1^+$ state in
    $^{48}$Ca.  The energy is calculated with SGII+(T,U)=(500,$-$280) and SAMi-T,(T,U)=(415.5,$-$91.4) without and with tensor terms.   $\Delta$ E$^T$ is the difference of energies between
    with and without the tensor interaction,    while $\Delta$E(RPA-HF) ($\Delta E$(SSRPA-HF) is the difference between RPA (SSRPA) and HF energies.   
    The experimental excitation energy is 10.23 MeV.  
    The unit is in  MeV. See the text for details.}
  \label{tab1} 
  \vspace{2mm}
    \begin{tabular}{cccccc}
      \hline
SGII       &  HF & RPA  & $\Delta$E(RPA-HF)   & SSRPA  & $\Delta E$(SSRPA-HF) \\ \hline
      w/o &  6.44 & 8.90 &  2.46 & 7.34 & 0.90 \\
      with &  9.25 & 11.35 & 2.10  & 9.05 & $-$0.20 \\ \hline
     $\Delta$ E$^T$ & 2.81 &  2.45 &    &  1.71  &      \\
           \hline
SAMi-T       &  HF & RPA  & $\Delta$E(RPA-HF)   & SSRPA  & $\Delta E$(SSRPA-HF) \\ \hline
      w/o &  6.18 & 8.61 &  2.43 & 7.88 & 1.70 \\
      with &  7.01 & 9.40& 2.39  & 8.34 & 1.33 \\ \hline
     $\Delta$ E$^T$ & 0.83 &  0.46 &    &  $-$0.37  &      \\\hline
  \end{tabular}
\end{table*}

\begin{figure*}[t]
\centering
\includegraphics[width=16cm,clip,bb=0 0 400 270]{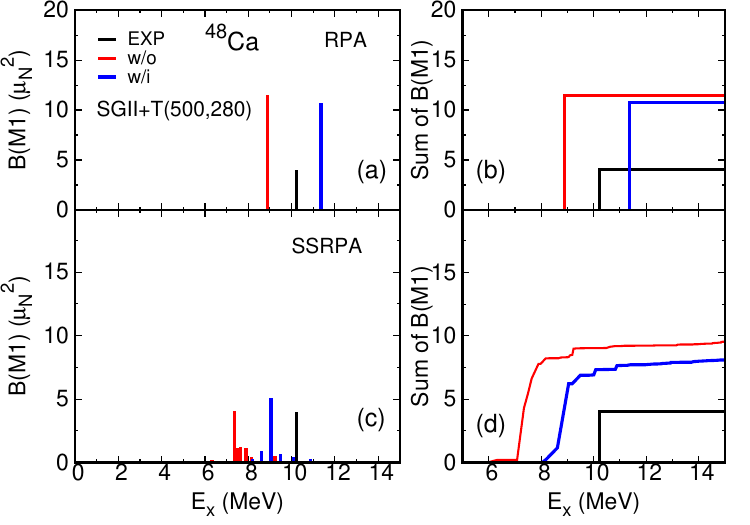} 
\caption{(left panels) M1strengths  of $^{48}$Ca  calculated by RPA and SSRPA models. The red and blue lines shows results of the SGII,  SGII+T(500,-280) EDFs without/with the tensor terms, respectively.  (right panels) Cumulated sum of the M1 strength.  
 Red and blue solid curves correspond to the results without/ with tensor terms.  Experimental data are shown by black line taken from Ref. \cite{J.Birkhan2016}.}
\label{fig:M1-SSRPA}       
\end{figure*}
 In the left panel of Fig.  \ref{fig:M1-SSRPA},  RPA and SSRPA results of M1strengths in $^{48}$Ca are shown without  and with tensor terms by red and blue lines, respectively.  In the upper panel, the tensor correlations shift the energy of 1$^+$ state by about 2 MeV.  In the lower panel, the M1 strengths are very much fragmented by the effect of 2p-2h configuration mixings. The cumulative sums of B(M1) in the right panel are counted up to $\rm E_{max}$=15 MeV.  The cumulated sum shows a significant quenching effect, more than 20\%, due to the couplings to 2p-2h configurations. The tensor terms give further quenching for the sum by 5\% for RPA and 15\% for SSRPA.  The magnetic dipole transitions of other closed-shell nuclei were also discussed in Ref. \cite{MJ2024}. 

\subsection{$\beta$-decay lifetime}
\begin{figure*}[t]
\centering
\includegraphics[width=17cm,clip,bb=0 0 900 520]{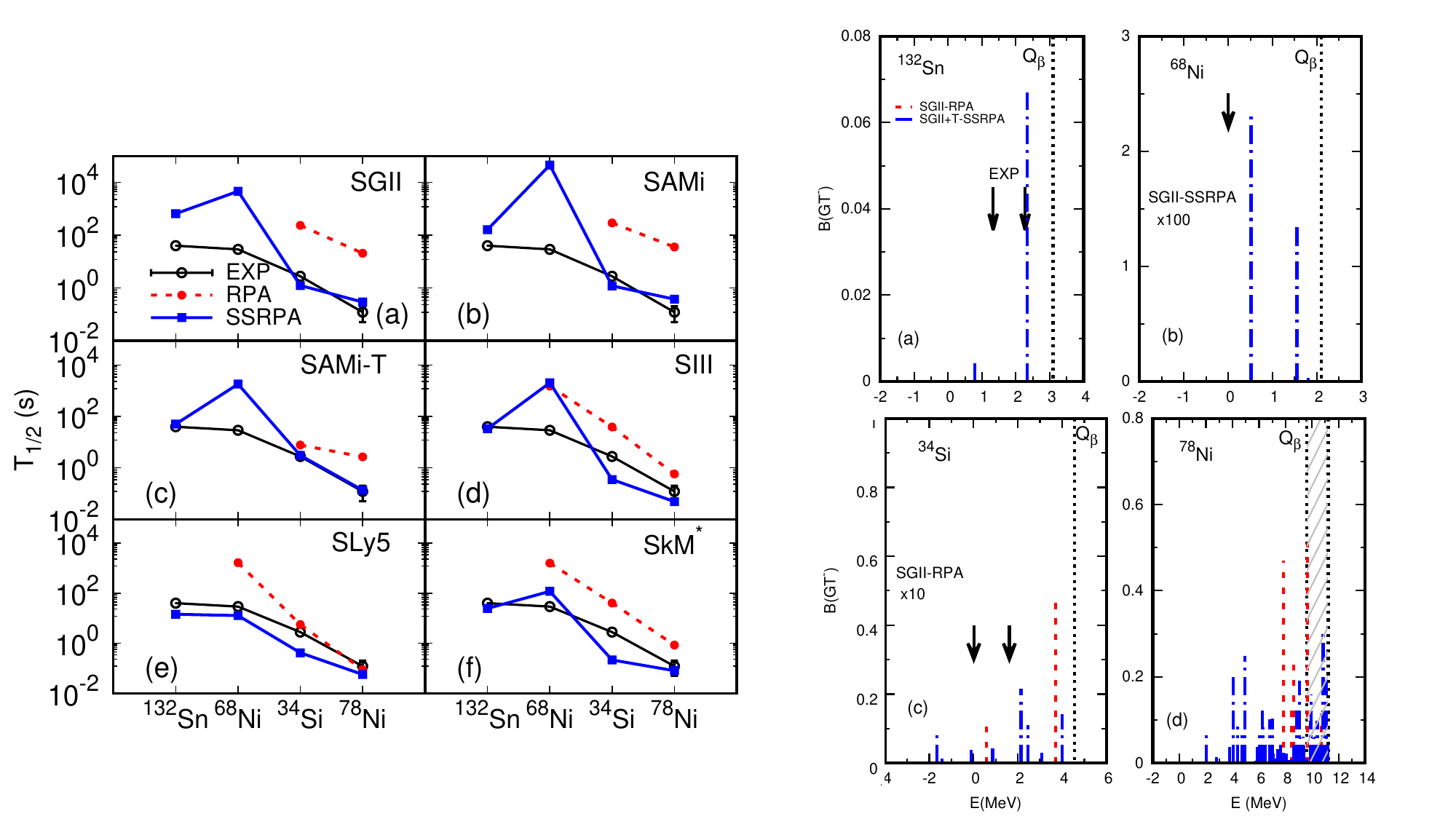} 
\caption{(left panels) $\beta$ decay life time calculated by RPA (red dashed line) and SSRPA (blue solid line) for several Skyrme EDFs.  The experimental data are shown by black line.   (right panels) B(GT) values blow $Q_\beta$ value.  RPA results are shown by red dashed lines, while SSRPA results are 
shown by blue dashed-dotted lines.  The arrows show energies of observed $1^+$ states.  Experimental data are  taken from Ref. \cite{NNDC}. The RPA results are infinite in some nuclei and	are not shown in the figure.}
\label{fig:beta-decay}       
\end{figure*}

The GT-type $\beta$ decay half-life
can be calculated using a formula~\cite{J.Engel1999}:
\begin{equation}\label{eq3}
   T_{1/2}=\frac{D}{g_A^2\sum_n  B_{1^+_n }^{GT_-}f_0(Z,A,\omega_n ) },   
\end{equation}
where $D=6163.4 \pm 3.8$ s, $g_A\equiv G_A /G_V= 1.26$ is the 
ratio of the axial-vector to vector coupling constants, {$B_{1^+_n }^{GT_-}$ is the Gamow-Teller strength of $n$-th excited state with the energy $\omega_n $ referred 
to the ground state of mother nucleus. The Gamow-Teller operator is defined as}
\begin{equation} 
\hat{O}(GT_-)=\sum_i t_-(i){\bm \sigma}(i).  
\end{equation} 
 The factor $f_0(Z,A,\omega_n )$ is 
the integrated phase factor at the energy $\omega_n$. 
The value $g_A$ is usually set to lower than 1.26 assuming a quenching factor which is closely related to 
the GT sum rule deficiency ~\cite{E.Caurier2005}.
In this work, the value $g_A$ is set to be $g_A=1.0$ \cite{MJ2024}.  This value is
consistent to the quenching factor in our previous work 
on the study of  GT transition strengths 
by SSRPA model~\cite{MJ2022}.  
We study the effect of the 2p-2h correlations 
on the $\beta$-decay half-live of the four semi-magic and magic nuclei $^{132}\rm Sn$, $^{68}\rm Ni$, 
$^{34}\rm Si$, and $^{78}\rm Ni$.

Figure~\ref{fig:beta-decay} shows the $\beta$ decay half-lives calculated by RPA and SSRPA models, 
in comparisons with experimental values. The RPA results
largely  overestimate the half-lives for almost all
nuclei. For example,  $^{132}\rm Sn$ has infinite lifetime   in RPA calculations of all EDFs. 
On the other hand, the half-lives of all nuclei calculated with SSRPA become  finite values, 
 and agreements with  the experimental values are much improved.  
In Fig.~\ref{fig:beta-decay}, the SSRPA results of SLy5 and $\rm SkM^*$ EDFs give 
 better agreements with the experimental half-lives 
   than the other EFDs. 
Similar improvements of $\beta$ decay descriptions  were obtained by
RPA+PVC calculations~\cite{Y.F.Niu2015}.

\section{Summary}
  We first discuss the analogy  between the subtraction procedure of SSRPA matrices and the similarity transformation by Okubo-Lee-Suzuki method for beyond mean field calculations.   Namely,  the subtracted matrix elements of SSRPA from the $1p$-$1h$ model space corresponds 
   to the renormalization terms from $Q$ space to the effective interaction  of $P$ space. This similarity may justify the subtraction procedure  of SSRPA. 
    We applied the SSRPA model to describe the magnetic dipole states,  GT states and $\beta$ decays.   It is pointed out the quenching of M1 and GT states are largely explained by the couplings to $2p$-$2h$ states. On top of that,   the tensor interaction contributes 5-10 \% effect to induce further  the quenching effect.    The $\beta$ decay life-times of semi-magic and magic nuclei were  also studied in the SSRPA model and a large improvement of the lifetime predictions is  found in these nuclei compared with the standard RPA calculations.
Thus, the higher order effects more than the RPA are indispensably needed for the quantitative discussions of spin-isospin excitations.  

\textit{Acknowledgments}

I acknowledge long-term collaboration on beyond mean field model project with Mingjun Yang and Chunlin Bai.  
I   also  thank Eunja Ha,   Myung-Ki Cheoun,  Shangui Zhou, Ligang Cao and Gianluca Col\`o for fruitful discussions on microscopic models of nuclear structure calculations.
This work is supported by Chinese Academy of Sciences CAS President's International Fellowship Initiative (PIFI) Grant No. 2024PVA0003-Y1.

\end{document}